\documentclass[a4paper,superscriptaddress,floatfix]{revtex4}

\usepackage{graphicx,multirow}
\usepackage{bm}
\usepackage{amsmath}
\usepackage{amssymb}
\usepackage{amscd}
\usepackage{latexsym}
\usepackage{slashed}
\usepackage{color}
\usepackage{graphicx}
\usepackage{ulem}
\usepackage{color}
\usepackage[caption=false]{subfig}

\def\ltap{\raisebox{-.6ex}{\rlap{$\,\sim\,$}} \raisebox{.4ex}{$\,<\,$}}

\begin{document}

\begin{flushright}
%\hspace{3cm} 
MS-TP-19-30\\
Cavendish-HEP-19/16\\
\end{flushright}

\title{
{Production of $Z^\prime$-Boson Resonances with Large Width at the LHC}
}

\author{E. Accomando}%
 \email{e.accomando@soton.ac.uk}
\affiliation{School of Physics and Astronomy, University of Southampton, Highfield, Southampton SO17 1BJ, UK}%
\author{F. Coradeschi}% 
 \email{f.coradeschi@damtp.cam.ac.uk}
\affiliation{DAMTP, CMS, University of Cambridge, Wilberforce road, Cambridge, CB3 0WA, UK}%
\author{T. Cridge}%
 \email{t.cridge@ucl.ac.uk}
 \affiliation{Department of Physics and Astronomy, University College London, WC1E 6BT, UK}%
\author{J. Fiaschi}%
 \email{fiaschi@uni-muenster.de}
 \affiliation{Institut f\"ur Theoretische Physik, Universit\"at M\"unster, D 48149 M\"unster, Germany}%
\author{F.~Hautmann}%
 \email{hautmann@thphys.ox.ac.uk}
\affiliation{Elementaire Deeltjes Fysica, Universiteit Antwerpen, B 2020 Antwerpen, Belgium}
\affiliation{Theoretical Physics Department, University of Oxford, Oxford OX1 3NP, UK}%
\affiliation{UPV/EHU University of the Basque Country, Bilbao 48080}
\affiliation{CERN, Theoretical Physics Department, CH 1211 Geneva, Switzerland}
\author{S. Moretti}%
 \email{s.moretti@soton.ac.uk}
\affiliation{School of Physics and Astronomy, University of Southampton, Highfield, Southampton SO17 1BJ, UK}%
\author{C. Shepherd-Themistocleous}%
 \email{claire.shepherd@stfc.ac.uk}
\affiliation{Particle Physics Department,
STFC, Rutherford Appleton Laboratory, Harwell Science and Innovation Campus,
Didcot, Oxfordshire, OX11 0QX, UK}%
\author{C. Voisey}
 \email{voisey@hep.phy.cam.ac.uk}
\affiliation{Cavendish Laboratory, University of Cambridge, Cambridge CB3 0HE, UK}

\begin{abstract}
\noindent
Di-lepton searches for Beyond the Standard Model (BSM) $Z'$ bosons that
rely on the analysis of the Breit-Wigner (BW) line shape are
appropriate in the case of narrow resonances, but likely not
sufficient in scenarios featuring $Z^\prime$ states with large
widths. Conversely, alternative experimental strategies
applicable to wide $Z^\prime$ resonances are much more dependent than
the default bump search analyses on the modelling of QCD higher-order
corrections to the production processes,  for both signal and background. 
For heavy $Z'$ boson searches in the di-lepton
channel at the CERN Large Hadron Collider (LHC), the transverse
momentum $q_T$ of the di-lepton system peaks at $q_T \ltap 10^{-2}
M_{ll}$, where $M_{ll}$ is the di-lepton invariant
mass. We exploit  this to treat the QCD corrections  by using the
logarithmic resummation methods in $M_{ll} / q_T$ to all orders
in the strong coupling constant $\alpha_s$. We carry out studies of
$Z^\prime$ states with large width at the LHC by employing the program
{\tt reSolve}, which performs QCD transverse momentum resummation up
to Next-to-Next-to-Leading Logarithmic (NNLL) accuracy. We consider
two benchmark BSM scenarios, based on the Sequential Standard Model
(SSM) and dubbed `SSM wide' and `SSM enhanced'. We present results for
the shape and size of $Z'$ boson signals at the differential
level, mapped in both  cross section ($\sigma$) and Forward-Backward Asymmetry ($A_{\rm FB}$), and
perform numerical investigations of the experimental sensitivity at
the LHC Run 3 and  High-Luminosity LHC (HL-LHC).
\end{abstract}

\maketitle

\section{Introduction}

\noindent

The physics of $Z'$ bosons has been extensively studied in the
literature. For an exhaustive review, e.g.,  see
Ref.~\cite{Langacker:2008yv}. There are numerous BSM scenarios in
which the predicted $Z'$ boson is characterised by a large
width $\Gamma_{Z^\prime}$. Examples of 
these include Technicolour \cite{Belyaev:2008yj}
and Composite Higgs Models 
\cite{Barducci:2012kk,Accomando:2015cva,Agashe:2004rs,DeCurtis:2011yx}, where additional $Z'$ boson decay channels into exotic
particles can take place. Model configurations also exist where the
$Z'$ boson generally couples to the first two generations and the third 
one differently \cite{Kim:2014afa, Malkawi:1999sa}. It is
notable that the couplings to the latter are not constrained by the
most stringent $Z^\prime$ searches, i.e., those in Drell-Yan (DY)
di-electron and di-muon
channels~\cite{Basso:2012ux,Basso:2012sz,Barducci:2012sk,Accomando:2013dia,Cerrito:2016bvl,Cerrito:2016xga}.
Large $\Gamma_{Z^\prime}/M_{Z^\prime}$ values can result from such
phenomena. A  wide $Z^\prime$ resonance does not have an
easily observable narrow BW line shape, instead it appears as a broad
shoulder spreading over the SM background.  In the above
circumstances, the ratio $\Gamma_{Z^\prime}/M_{Z^\prime}$ can easily
reach a magnitude of 50$\%$, making a classical narrow BW line shape
based analysis inappropriate.

Alternative experimental approaches can be applied to the case of a
wide $Z^\prime$ resonance. Non-resonant searches, such as simply
counting the number of events appearing above a certain lower
threshold in the di-lepton invariant mass spectrum and comparing this
measured value with the theoretical SM expectation, are for example 
performed. Another approach is to make use of additional observables
supporting and/or complementing $\sigma$ mapped in
the di-lepton invariant mass $M_{ll}$. 
% (also denoted as $m_{ll}$).
In this  context, a simple observable that has been shown to be quite
effective is $A_{\rm
  FB}$~\cite{Accomando:2015sun,Accomando:2015cfa,Accomando:2015ava,Accomando:2015pqa,Accomando:2016mvz,Accomando:2017fmb}.

Counting strategies rely heavily on the knowledge of the SM background
in the large invariant mass region. Experiments generally use a
combination of information from Monte Carlo (MC) simulation and data to
estimate the SM background in the high-mass region of interest. An
example approach is to parameterise a functional form using simulation
and then constrain the overall amplitude using a low-mass control
region assumed to be free from significant new physics content. This
then provides a background estimate in the signal region of interest.
The quality of this estimate will be subject to systematic
uncertainties due to the theoretical understanding of the background,
such as those due to Parton Distribution Functions (PDFs) or to missing
higher order terms in theoretical calculations. As a result, having QCD
corrections to the di-lepton spectrum well under control is
of significant importance. Since $A_{\rm FB}$ is a ratio
quantity some systematics will cancel and it is expected that QCD
higher-order corrections will be lower than in the case of cross
sections (and so their residual systematics).  This was established to be the case for the PDF error in
Ref.~\cite{Accomando:2015cfa}, in fact, to the extent that the
differential $A_{\rm FB}$ in combination with the differential
$\sigma$ can even be used to improve PDF fits over a wide di-lepton
invariant mass range of neutral DY final states
\cite{Accomando:2017scx,Fiaschi:2018buk,Accomando:2018nig,Fiaschi:2019skn,Abdolmaleki:2019qmq,Abdolmaleki:2019ubu}. As the shape of $d\sigma/dM_{ll}$ and $dA_{\rm  FB}/dM_{ll}$
can change by factors that do not correspond to a
trivial overall rescaling factor, one has to account for such QCD effects in
the relevant experimental searches. In fact, it is of
crucial importance to determine the impact of such corrections not
only in these two differential distributions but also on the variables
used for the selection of neutral DY events, i.e., the individual
lepton transverse momentum ($p_T^l$) and  pseudorapidity
($\eta_l$), as these may also be affected non-trivially.

This paper is motivated by the observation that, in the multi-TeV mass
range of $Z'$ boson searches, the transverse momentum $q_T$
of the di-lepton system peaks at values much smaller than its
invariant mass, $q_T \ltap 10^{-2} M_{ll}$.  We thus exploit the
fact that the peaks in the $q_T$ and $M_{ll}$ distributions are
two orders of magnitude apart to treat QCD higher-order corrections by
using resummation
techniques~\cite{Dokshitzer:1978hw,Parisi:1979se,Curci:1979bg,Collins:1981uk,Kodaira:1981nh,Collins:1984kg,Catani:1988vd,deFlorian:2000pr,Catani:2010pd,Catani:2013tia}. These
techniques take into account logarithmically-enhanced
contributions $\alpha_s^k \ln^n (M_{l l}/ q_T)$ ($n \leq 2 k$)
to the differential cross section to all orders in $\alpha_s$ with
NNLL accuracy, and neglect power suppressed contributions of order
${\cal O} (q_T / M_{l l})$. A number of computational tools
implementing this method are
available~\cite{Bozzi:2010xn,Catani:2015vma,Camarda:2019zyx,Coradeschi:2017zzw,Hautmann:2017xtx,Hautmann:2017fcj,Martinez:2018jxt,Martinez:2019mwt,Bizon:2018foh,Bizon:2019zgf,Scimemi:2017etj,Scimemi:2018xaf,Ebert:2016gcn,Ebert:2017uel,Billis:2019vxg,Becher:2011xn,Becher:2012yn,Landry:2002ix,Konychev:2005iy}. Currently,
a benchmarking exercise based on these tools is being performed within
the LHC Electroweak Working Group \cite{LHCworkingroup:2019} in the
context of precision DY studies in the SM. (Further information may be found in
Ref.~\cite{bozzi:2019}.) In this paper, we use the code {\tt
  reSolve} described in \cite{Coradeschi:2017zzw}, modified for our
purposes to include the $Z'$ boson contribution to the DY-channel
in addition to the default SM ones ($\gamma,Z$).

The plan of this paper is as follows. In Sec.~2 we describe two
illustrative theoretical frameworks embedding a $Z'$ boson
of significant width, which are called `SSM wide' and `SSM enhanced',
where each is a variant of the SSM scenario
\cite{Altarelli:1989ff}. In Sec.~3 we present the results, and give
conclusions in Sec.~4.

\section{Benchmark models}

In this section, we introduce the benchmark models that we use to
implement $Z^\prime$ resonances with large width. The model taken to
be a reference model by the LHC experimental collaborations, ATLAS and
CMS, when searching for wide $Z'$ bosons, is the
SSM~\cite{Altarelli:1989ff} with the standard couplings given
in~\cite{Accomando:2010fz}. We consider two variants of this model,
which we call `SSM wide' and `SSM enhanced'.

In the SSM wide variant, the resonance width is enhanced by the
opening of extra invisible decay channels. The chiral couplings to
ordinary matter are unchanged with respect to the usual SSM model,
while the resonance width is modified so as to have $\Gamma_{Z^\prime}
/ M_{Z^\prime} = 10\%$.  In this scenario the resonance peak still has
a BW line shape but it is broad enough to possibly escape detection
via the standard bump hunt method. In the SSM enhanced case, a  different
mechanism is used and in this case the resonance width is enhanced by
increasing the coupling between the $Z'$ boson and the ordinary
matter. We rescale the couplings by a factor of three with respect to the
usual SSM model.

\begin{table}[t]
\begin{center}
\begin{tabular}{|c|c|c|c|c|c|c|c|c|c|c|c|}
\hline
$U(1)^\prime$ & $M_{Z^\prime}({\rm GeV})$ & $\Gamma_{Z^\prime}/M_{Z^\prime}$ & 
$g^\prime$ & $g_V^u$ & $g_A^u$ & $g_V^d$ & $g_A^d$ & $g_V^e$ & $g_A^e$ & $g_V^\nu$ & $g_A^\nu$ \\  
\hline
$U(1)_{\rm SSM wide}$ & 4500 & 10$\%$ & 0.76 & 0.193 & 0.5 & $-0.347$ & $-0.5$ & $-0.0387$ & $-0.5$ & 0.5 & 0.5 \\	
\hline
$U(1)_{\rm SSM enhanced}$ & 5000 & 27$\%$ & 2.28 & 0.193 & 0.5 & $-0.347$ & $-0.5$ & $-0.0387$ 
        & $-0.5$ & 0.5 & 0.5 \\
\hline
\end{tabular}
\caption{
Mass, width-to-mass ratio and couplings to fermions of the $Z'$ boson in the SSM wide and SSM enhanced benchmark scenarios.}
\label{tab:parameters}
\end{center}
\end{table}

We set the model parameters so that the $Z^\prime$ resonance is beyond
the current sensitivity of the ATLAS~\cite{Aad:2019fac} and
CMS~\cite{Sirunyan:2018exx, CMS:2018wsn}  experiments at the LHC. In order to
extract the bounds on the $Z'$ boson mass and couplings, we use
the computer codes of Refs.~\cite{Accomando:2013sfa,
  Accomando:2015cfa}. There, the significance of the BSM signal is
estimated from the partonic cross section evaluated at Leading Order (LO) and
convoluted with CT14NNLO PDF~\cite{Dulat:2015mca}, with the addition
of a mass dependent $k$-factor accounting for Next-to-LO (NLO) and Next-to-NLO (NNLO) QCD
corrections~\cite{Hamberg:1990np, Harlander:2002wh} and
acceptance as well as efficiency factors for both the di-electron and di-muon
final states given in Ref.~\cite{Khachatryan:2014fba}. We then combine
the significance for each of the two di-lepton channels to obtain the
overall significance of the $Z^\prime$ resonance. In this way, we
estimate the latest limits coming from the LHC Run 2 data analysis and
set the benchmark model parameters accordingly.

Mass, width-to-mass ratio and couplings to the ordinary matter of the
$Z'$ boson within the two chosen benchmark scenarios are
summarised in Tab.~\ref{tab:parameters}.

%In order to overcome such technical difficulties, other extra observables are frequently included to boost the experimental sensitivity to wide $Z^\prime$s such as the (reconstructed) Forward-Backward Asymmetry ($A_{\rm{\rm{FB}}}^*$)~\cite{Accomando:2015cfa} and the transverse momentum $p_T$ spectrum~\cite{Accomando:2017fmb}.

\section{Numerical results}

In this section, we show the results obtained by using the MC
program {\tt reSolve}~\cite{Coradeschi:2017zzw}, which performs the described 
transverse momentum resummation.  This is a tool to compute
differential distributions for colourless final states in
hadron-hadron collisions, incorporating Born-level matrix element and
QCD transverse momentum
resummation~\cite{Dokshitzer:1978hw,Parisi:1979se,Curci:1979bg,Collins:1981uk,Kodaira:1981nh,Collins:1984kg,Catani:1988vd,deFlorian:2000pr,Catani:2010pd,Catani:2013tia}
up to NNLL accuracy. 
The initial public version, {\tt reSolve-1.0}~\cite{Coradeschi:2017zzw}, focussed purely on the Standard Model, however for the benefit of this analysis it
has been complemented with the addition of $Z'$ boson production and
decay, including interference with the SM $\gamma, Z$
contribution. A complete description of the new version of the code
will be given elsewhere~\cite{Coradeschi:2019}. Here, we just
summarise the main features.

The resummed calculation implemented in {\tt reSolve} is designed to
take into account QCD higher-order logarithmic corrections of the type
$\alpha_s^k \ln^n (M_{l l}/ q_T)$, $n \leq 2 k$, to the
differential cross section $ d \sigma / (d q_T d M_{l l} dY
d\Omega)$ (where $q_T$, $ M_{l l}$ and $Y$ are the di-lepton
transverse momentum, invariant mass and pseudorapidity, while $\Omega$
represents any additional variables internal to the final state that
may be needed to fully define its phase space) up to NNLL for any $k$,
by neglecting power-suppressed contributions of order ${\cal O} (q_T /
M_{l l})$ at each order $k \geq 1$. The use of this tool for the
calculations that follow is motivated by the fact that our search
window is in the large invariant mass region, $M_{l l}\ge 3 {\rm
  TeV}$, while the di-lepton transverse momentum distribution 
peaks around $q_T \sim {\cal O}$(10 GeV). The contributions from the
non-logarithmic high-$q_T$ tail of the distribution are thus expected
to be power-suppressed in the $Z^\prime$ relative to the SM case.
%This computation therefore includes all of the effects dominant at
%the low transverse momentum end of the spectrum, which we expect to
%extend to larger transverse momenta (compared to the SM) due to the
%large $Z'$ boson mass considered.
Given that the transverse momentum spectrum is strongly peaked at low
transverse momenta, the approximation adopted should account for the
majority of the contributions to the total cross section.
%The finite contribution to the cross section describing the presence of additional hard %jets beyond leading order, which are present and dominate at high transverse momentum, is %not included at this stage. Nonetheless the transverse momentum formalism implemented by %{\tt reSolve} does include virtual corrections, not involving the presence of an %additional hard jet, on top of the Born process through a hard factor \cite{Catani:%2013tia}. The last point to note is that {\tt reSolve} is a MC based tool, %therefore MC errors are associated with the calculated quantities, error bars on %Figures in this section will therefore indicate MC errors unless otherwise %stated.

\begin{figure}[!htbp]
\centerline{
  \subfloat{\includegraphics[height = 8cm, width = 8.5cm, trim= 2.5cm 0 0.8cm 2.5cm, clip,]{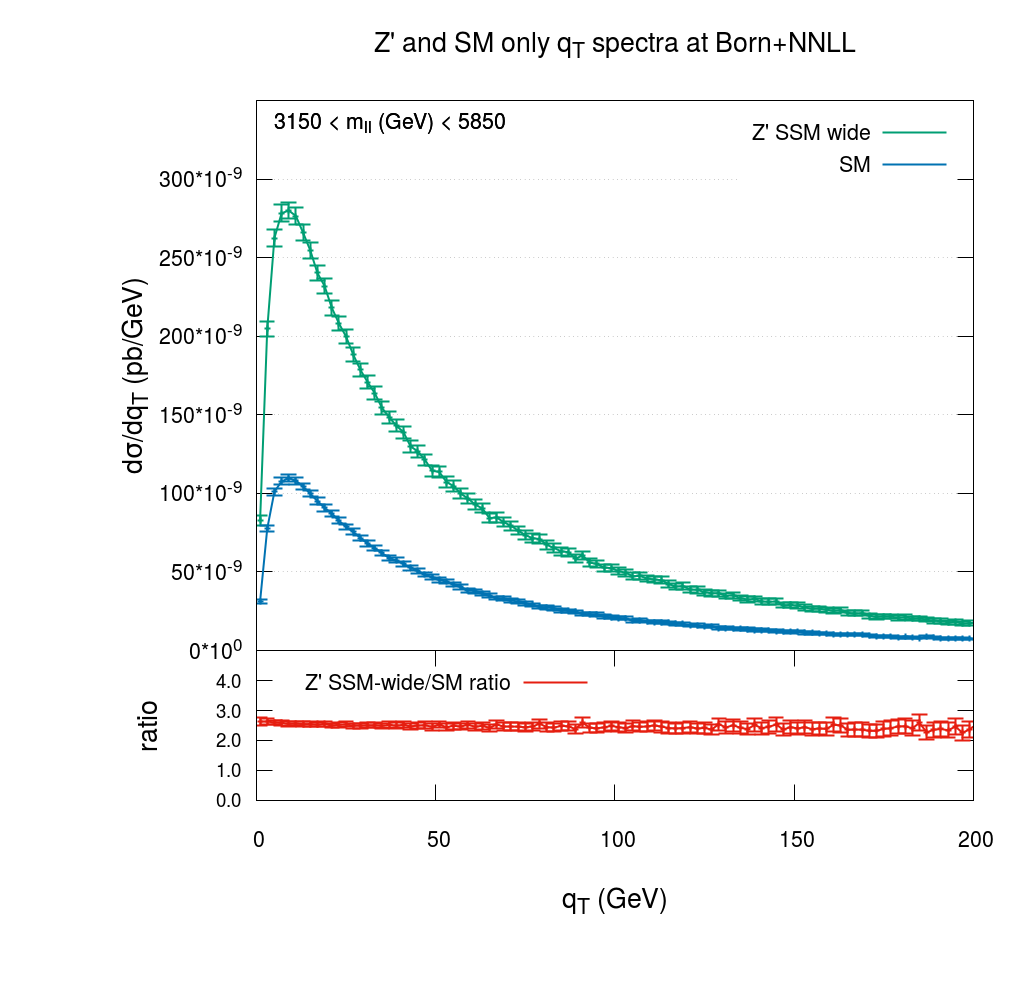}\label{qTspectrum_Z'vsSM_SSMwide} } 
  \subfloat{\includegraphics[height = 8cm, width = 8.5cm, trim= 2.5cm 0 0.8cm 2.5cm, clip,]{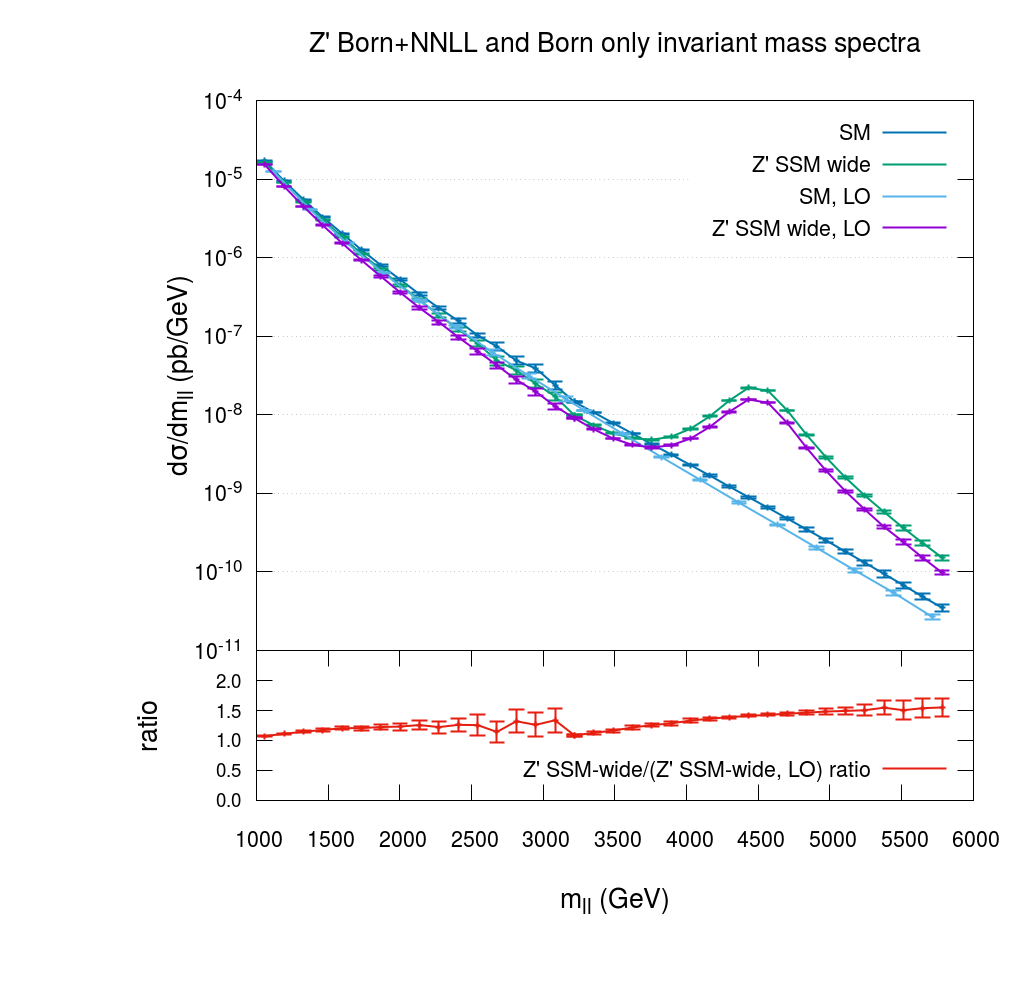}\label{qqspectrum_NNLLvsBorn_SSMwide} }
}\vspace{-0.8cm}
  \caption{\small{(a) Distribution in the transverse momentum of the di-lepton system calculated at NNLL for the SM (dark blue) and the SSM wide (green) scenario with 
$M_{Z^\prime} = 4.5 $ TeV. We select the invariant mass region relevant for the $Z'$ boson search $3150~{\rm GeV}\le M_{ll}\le 5850~{\rm GeV}$. The lower plot shows the ratio of the SSM wide $Z'$ boson $q_T$ spectrum to the SM one at NNLL. The bars indicate MC errors. (b) Di-lepton invariant mass ($M_{ll}$) distribution for the SM at LO (light blue) and NNLL (dark blue) and the SSM wide scenario at LO (purple) and at NNLL (green) with $M_{Z^\prime} = 4.5$ TeV. The lower plot shows the ratio between the NNLL and  LO invariant mass spectrum for the SSM wide case. The bars indicate MC errors.}}
\end{figure}

In Fig.~\ref{qTspectrum_Z'vsSM_SSMwide}, we show the distribution in the transverse momentum of the di-lepton system for the SM and the SSM wide scenario with $M_{Z^\prime} = 4.5$ TeV. We select the invariant masses in the range $3150~{\rm GeV}\le M_{l l} \le 5850~{\rm GeV}$, which corresponds to the search window $M_{Z^\prime} - 3 \Gamma_{Z^\prime}\le M_{l l} \le M_{Z^\prime} + 3 \Gamma_{Z^\prime}$ relevant to our studies. Clearly, the presence of the hypothetical $Z'$ boson causes a relatively uniform increase in the $q_T$ spectrum of about a factor of 3, compared to the SM values, at least in the low transverse momentum part dominated by the resummed contributions. Similar effects are seen for the SSM enhanced scenario. This is a pure NNLL prediction.  
%as, at the Born level, the di-lepton transverse momentum is null. 
The error bars shown in the plot are the MC errors. The statistical significance of the $q_T$ distribution will be discussed later for the two benchmark models, separately. Here, we just highlight the introductory features coming from the NNLL calculation, as compared to the Born result.
In Fig.~\ref{qqspectrum_NNLLvsBorn_SSMwide}, we show the effect of the
resummation on the di-lepton invariant mass spectrum in the SSM wide
scenario with the same mass as before, $M_{Z^\prime} = 4.5$ TeV. Here, we
see that higher-order corrections monotonically increase the
differential cross section, moving from low to high invariant masses,
reaching a roughly $50\%$ magnitude at the right end of the spectrum.

\begin{table}[t]
\begin{center}
\begin{tabular}{|c|c|c|c|}
\hline
	$U(1)^\prime$ & $p_T^1, p_T^2\ge 20~{\rm GeV}$ & $\eta_1, \eta_2 \le 2.5$ &  
        $p_T^1, p_T^2\ge 20~{\rm GeV}; \ \eta_1, \eta_2 \le 2.5$ \\
\hline
	$U(1)_{\rm SM}$ & 0.99 & 0.95 & 0.95 \\	
\hline
	$U(1)_{\rm SSM wide}$ & 0.99 & 0.96 & 0.96 \\
\hline
\end{tabular}
\caption{
Acceptances in the SM and SSM wide scenario for kinematical cuts in the transverse momentum and pseudorapidity of the single leptons. The mass window $3150~{\rm GeV}\le M_{ll}\le 5850~{\rm GeV}$ is selected. Here, the label 1(2) refers to the highest(lowest) transverse momentum lepton. }
\label{tab:acceptance}
\end{center}
\end{table}

%edits inthis para. ``Sensibly'' sentence changed.
Next, we consider the effect of a possible $Z'$ boson on the
acceptance by examining the distributions in the transverse momentum
and pseudorapidity of the individual leptons detected, at the NNLL
order. As before, we compare the SSM wide scenario with a
representative $Z'$ boson mass $M_{Z^\prime} = 4.5$ TeV to the SM
results. The presence of a wide $Z'$ boson produces a larger
differential cross section across all points in both differential
cross sections, as expected. In particular, the ratio BSM/SM has a
major increase at large absolute values of the lepton pseudorapidity and at
the Jacobian peak of the lepton transverse momentum distribution. The
acceptance however is not sensitive to higher-order
corrections. Tab. \ref{tab:acceptance} summarises the acceptance in
transverse momentum and pseudorapidity of the individual
leptons in the final state.

%\begin{figure}[!htbp]
%\centerline{
%  \subfloat{\includegraphics[height = 8cm,width=8.75cm,trim={2.9cm 0.6cm 0.1cm %2.6cm},clip]{Zp_SSM_wide_BornandNNLL_ZpvsSM_xsec_ratio_new_etaL.png}%\label{etaLspectrum_NNLL_ZpvsSM}}
%  \subfloat{\includegraphics[height = 8cm,width=8.75cm,trim={2.9cm 0.6cm 0.1cm %2.6cm},clip]{Zp_SSM_wide_BornandNNLL_ZpvsSM_xsec_ratio_new_pTmax.png}%\label{pTmaxspectrum_NNLL_ZpvsSM}}
%  \subfloat{\includegraphics[height = 8cm,width=8.75cm,trim={2.9cm 0.6cm 0.1cm %2.6cm},clip]{Zp_SSM_wide_BornandNNLL_ZpvsSM_xsec_ratio_new_pTmin_zoomed.png}%\label{pTminspectrum_NNLL_ZpvsSM}}
%}\vspace{-0.8cm}
%  \caption{\small{(a) Distribution in the rapidity of one of the leptons produced, calculated at NNLO. (b) Distribution in the minimum transverse momentum of one of the leptons in the final state, calculated at NNLO. In both cases, the SSM wide scenario with $M_{Z^\prime} = 4.5 $ TeV is compared to the SM. The ratio BSM/SM is shown in the lower subplots and error bars indicate MC errors only.}} %\label{trigger_efficiency_spectra}
%\end{figure} 

Then, we consider the discovery potential of the LHC for the wide $Z'$ boson predicted by the SSM wide and SSM enhanced scenarios, by employing two different techniques: the invariant mass event counting and the associated measurements of the spectrum and of three other support variables, i.e., $A_{\rm{FB}}$, the minimum transverse momentum of a single lepton, $p_T^{\rm min}$, and the transverse momentum of the di-lepton system, $q_T$. For these calculations, a 9-point scale variation was performed in {\tt reSolve} varying the renormalisation and factorisation scales by a factor of 2 around their default values and also, independently of this, varying the resummation scale by a factor of 2 up and down. As a result, we found that the scale dependence is negligible in comparison to the statistical errors for the cases we consider. We therefore omit its error band, as well as the MC errors (indeed sub-dominant), in the upcoming figures. Statistical error bands only are shown and only for the NNLL cases.

\begin{figure}[t]
\begin{center}
\includegraphics[width=0.5\textwidth]{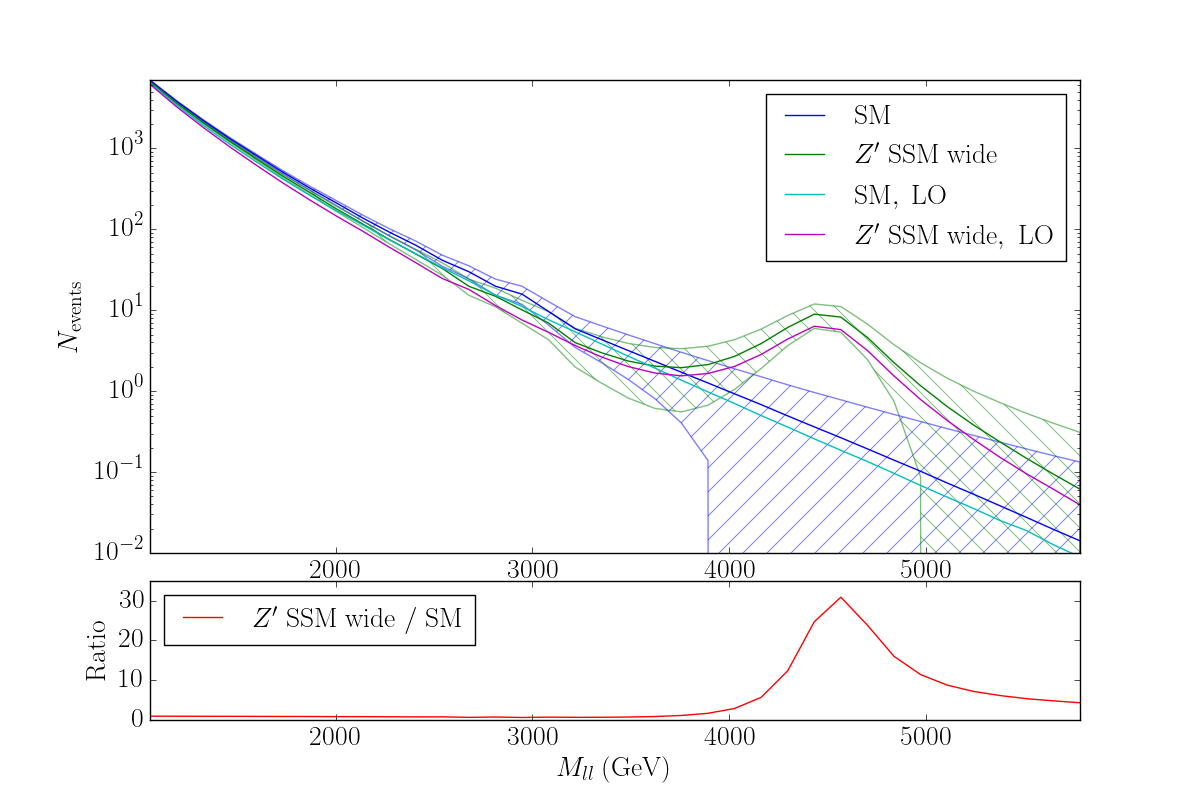}%{(a)}
\includegraphics[width=0.5\textwidth]{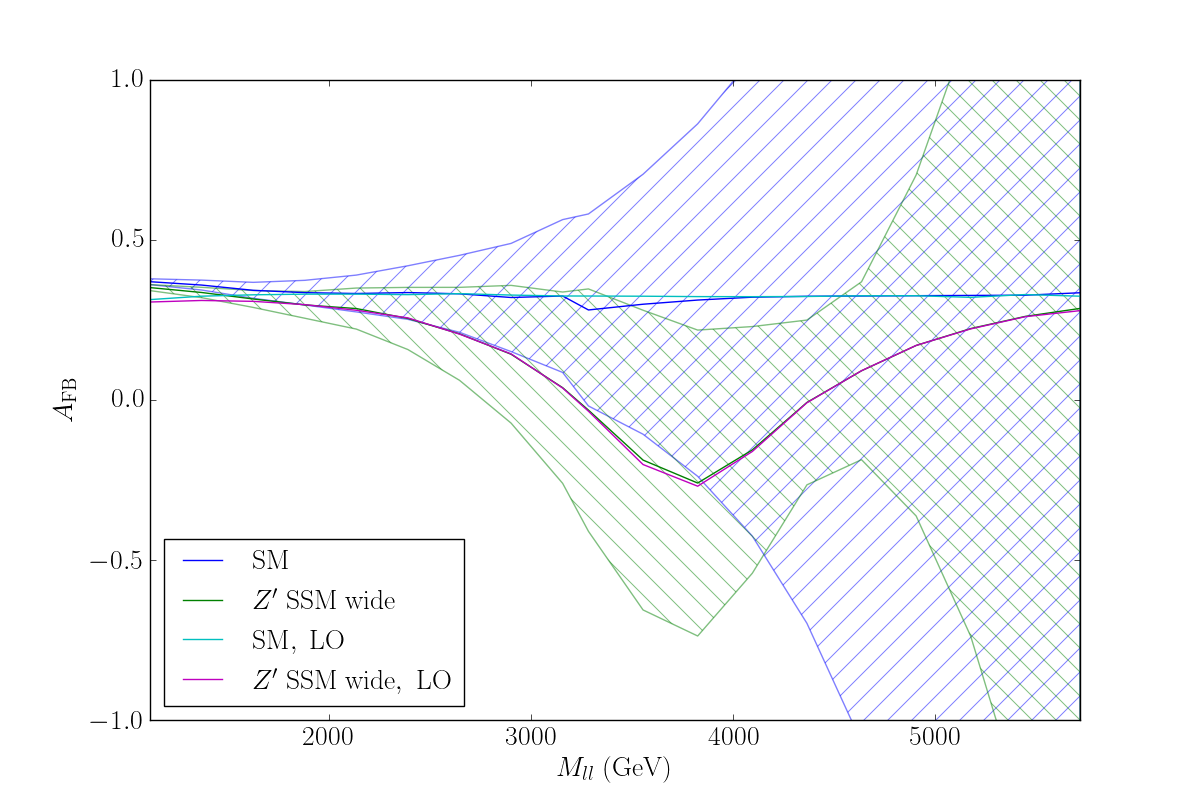}\\%{(b)}\\
\includegraphics[width=0.5\textwidth]{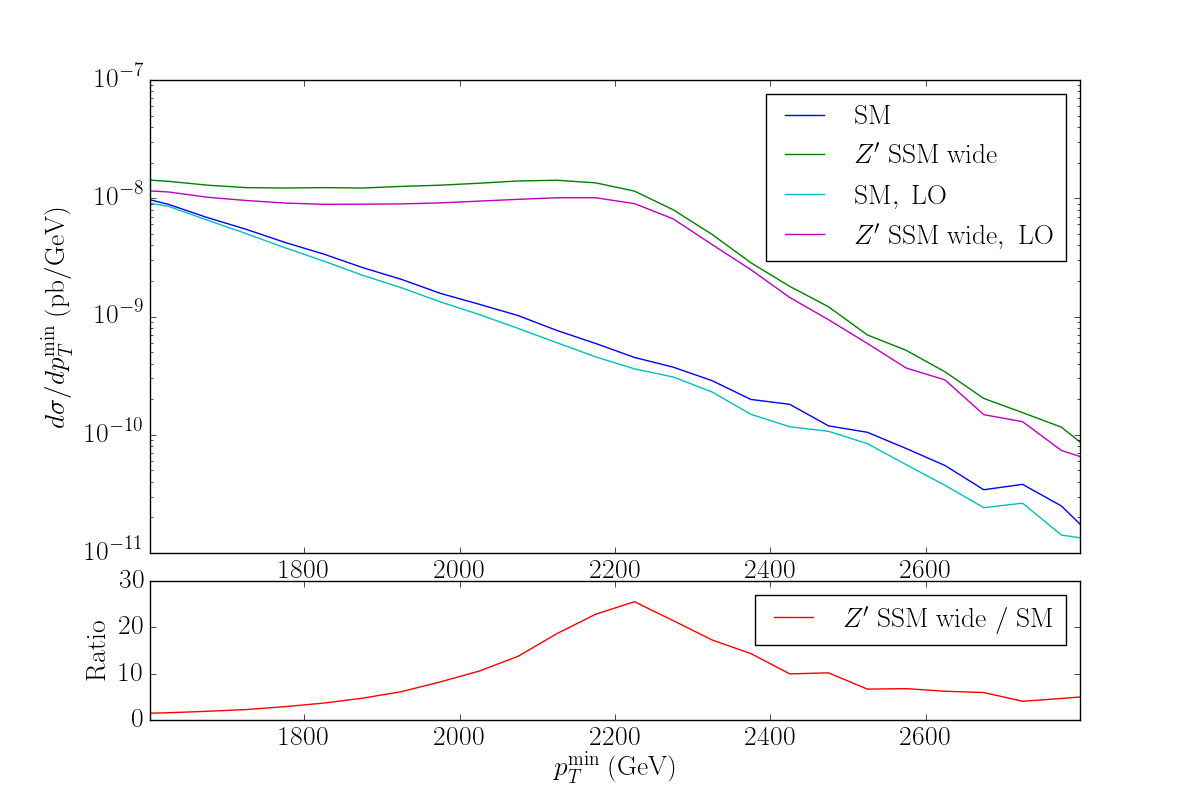}%{(c)}
\includegraphics[width=0.5\textwidth]{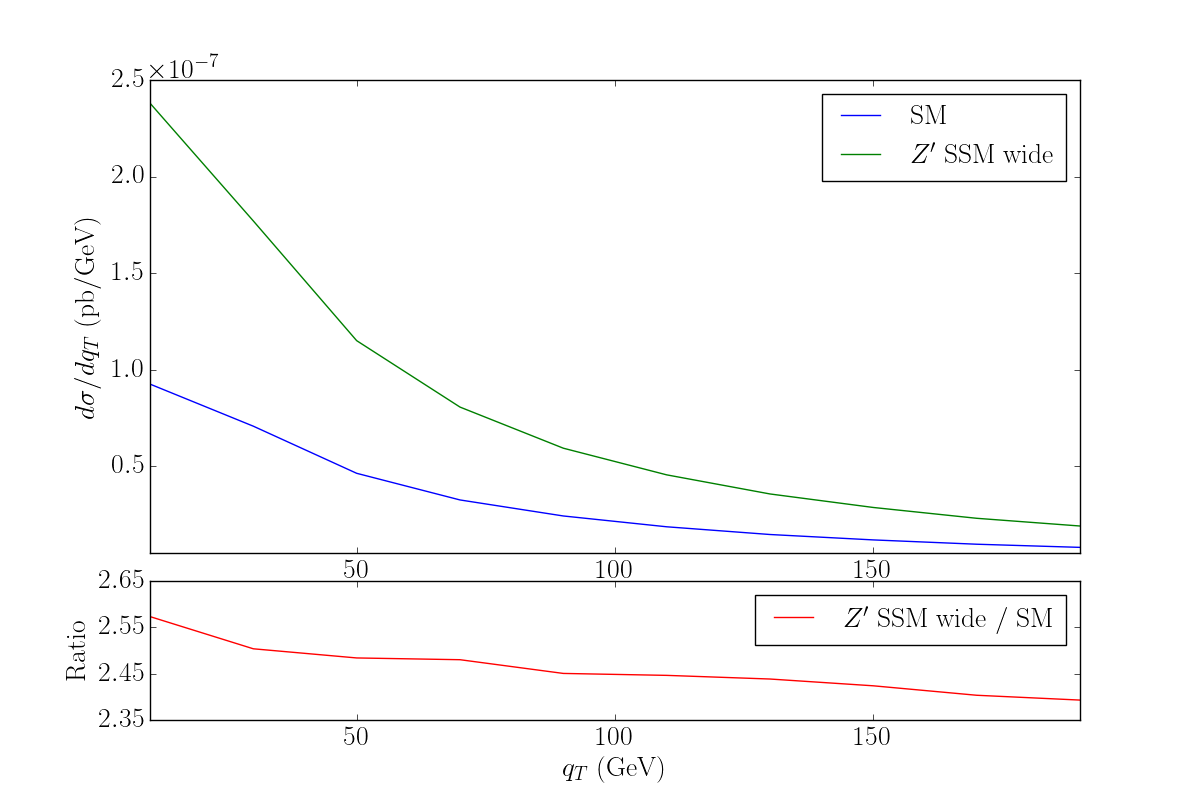}%{(d)}
\caption{\small{(a) Number of events versus the di-lepton invariant mass within the SM and the SSM wide scenario with $M_{Z^\prime}=4.5$ TeV. We assume an integrated luminosity $L = 3000$ fb$^{-1}$. The results are calculated at Born and NNLL order, along with the statistical error (dominant). The BSM/SM ratio is presented in the sub-plot. (b) Same as above for $A_{\rm FB}$  as a function of the di-lepton invariant mass. (c) Distribution in the minimum transverse momentum of a single lepton within the SM and the SSM wide scenario with $M_{Z^\prime}=4.5$ TeV, computed at the LO and NNLL order. The BSM/SM ratio is presented in the sub-plot. We select the invariant mass window $3150 ~{\rm GeV}\le M_{ll}\le 5850 ~{\rm GeV}$. (d) Same as in (c) for the distribution in the transverse momentum of the di-lepton system at NNLL. For all plots, acceptance cuts are applied (see fourth column in Tab.~\ref{tab:acceptance}).}}
\label{fig:numerics-SSM wide}
\end{center}
\end{figure}

We first discuss the SSM wide scenario. For this case, we consider an integrated luminosity of $3000$ fb$^{-1}$ corresponding to the HL-LHC. Fig.~\ref{fig:numerics-SSM wide}a presents the number of events versus the di-lepton invariant mass within the SM and the SSM wide scenario, calculated at both LO and NNLL. The error bands are purely statistical, as they completely dominate over the MC errors from our calculation with {\tt reSolve} and over the scale dependence. The $Z^\prime$ resonance peak at $M_{Z^\prime}=4.5$ TeV is clear and outside of the statistical errors, indicating that this model could be detected at the HL-LHC through the di-lepton event counting analysis.

$A_{\rm FB}$  for the same scenario is presented in Fig.~\ref{fig:numerics-SSM wide}b. As before, we show both the SM and the SSM wide scenario at LO and NNLL order. The large error bands represent the statistical uncertainty on the $A_{\rm{FB}}$ as calculated according to \cite{Accomando:2015ava}. The stability of the $A_{\rm FB}$ to 
 the higher orders is demonstrated, while the complementarity of the $A_{\rm{FB}}$ to the invariant mass spectrum is also clear, with the $A_{\rm{FB}}$ in the SSM wide scenario deviating from the SM at lower invariant masses and peaking well before the $Z'$ boson on-shell mass, $M_{Z^\prime}=4.5$ TeV. However, in this case the larger statistical uncertainty associated with $A_{\rm{FB}}$ suggests that the event counting analysis would offer greater promise. Any evidence in this spectrum could be made stronger by the additional measurement of the minimum transverse momentum of a single lepton, $p_T^{\rm min}$, and the di-lepton transverse momentum, $q_T$, given in Fig.s~\ref{fig:numerics-SSM wide}c and \ref{fig:numerics-SSM wide}d.
 
\begin{figure}[t]
\begin{center}
\includegraphics[width=0.5\textwidth]{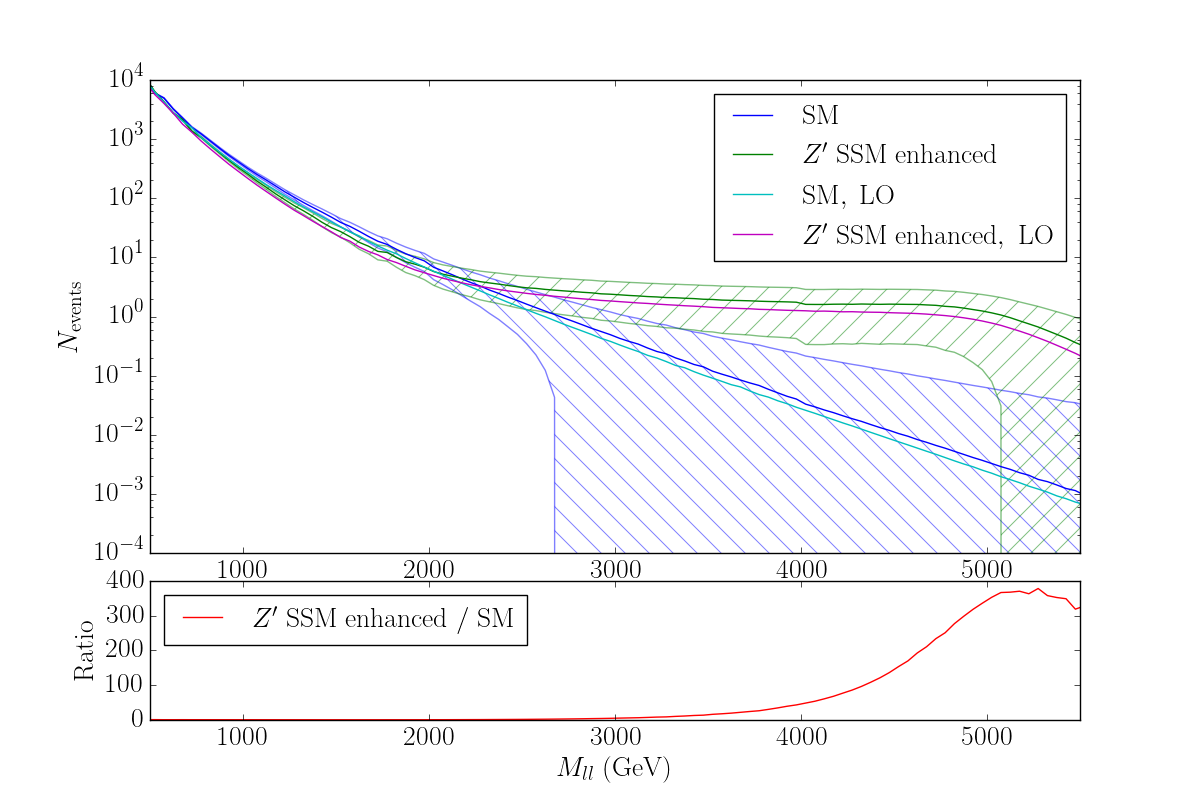}%{(a)}
\includegraphics[width=0.5\textwidth]{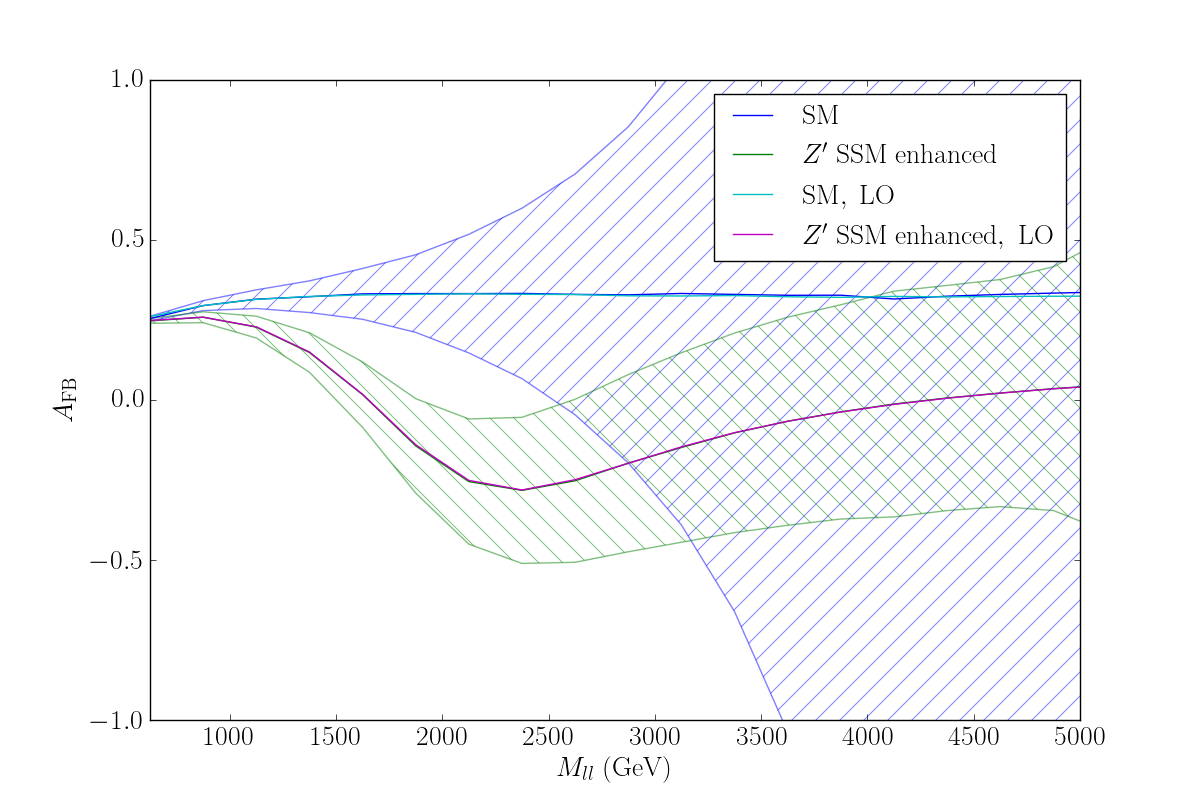}\\%{(b)}\\
\includegraphics[width=0.5\textwidth]{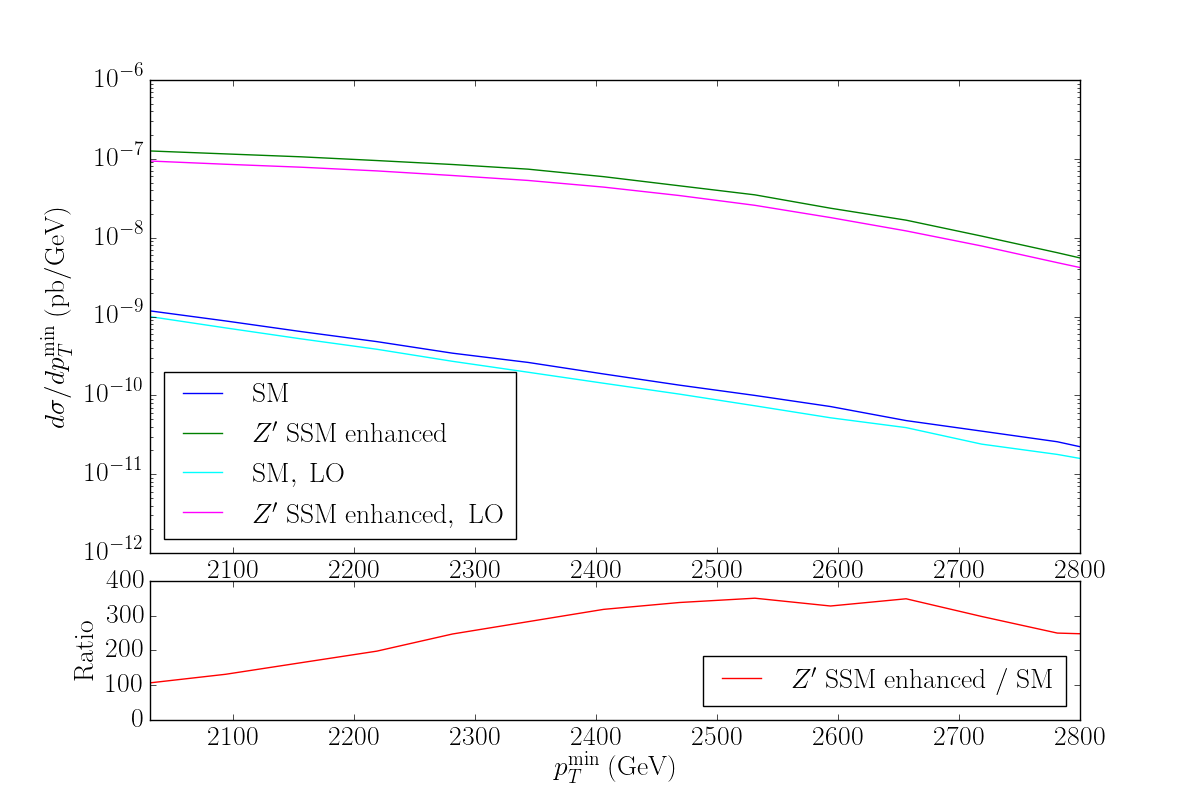}%{(c)}
\includegraphics[width=0.5\textwidth]{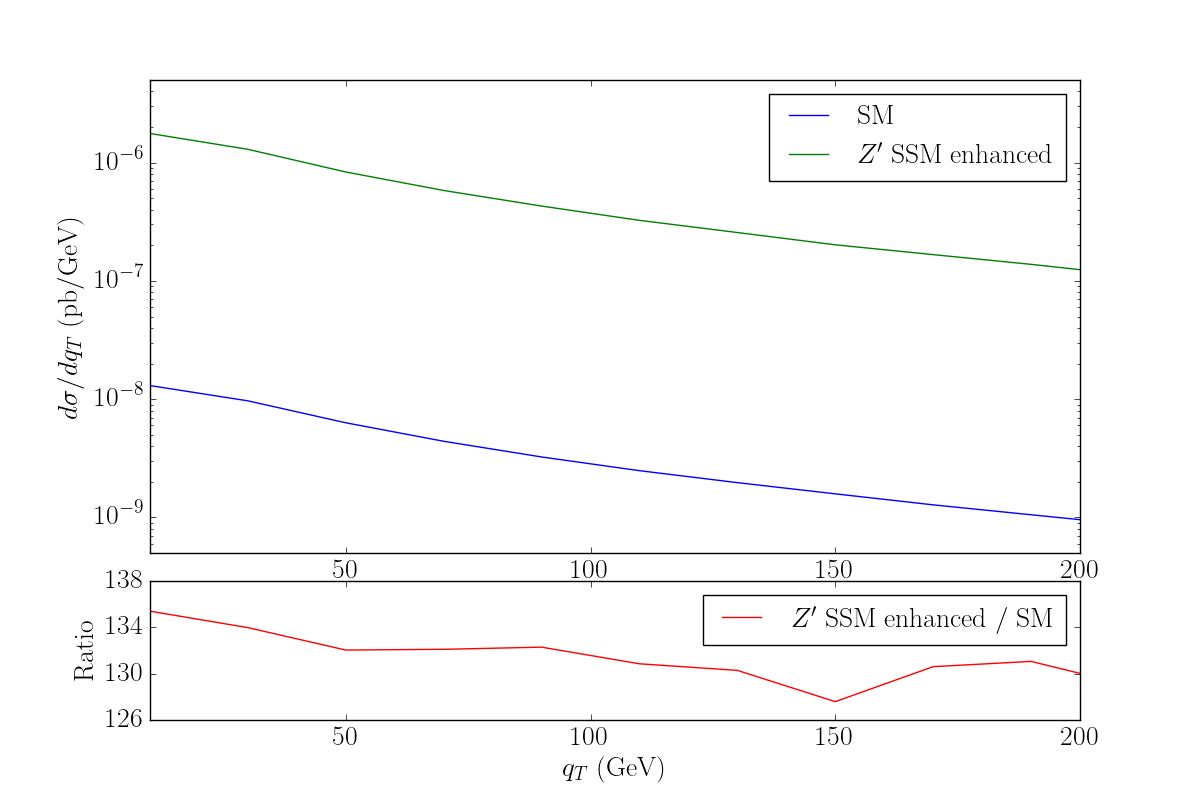}%{(d)}
\caption{\small{(a) Number of events versus the di-lepton invariant mass within the SM and the SSM enhanced scenario with $M_{Z^\prime}=5$ TeV. We assume an integrated luminosity $L = 300$ fb$^{-1}$. The results are calculated at the LO and NNLL orders, along with the statistical error (dominant) on the NNLL results. The BSM/SM ratio is presented in the sub-plot. (b) Same as above for $A_{\rm FB}$ as a function of the di-lepton invariant mass. (c) Differential cross section in the minimum transverse momentum of a single lepton within the SM and the SSM enhanced scenario with $M_{Z^\prime}=5$ TeV, computed at the LO and NNLL orders. The BSM/SM ratio is presented in the sub-plot. We select the invariant mass window $4000 ~{\rm GeV}\le M_{ll}\le 6000 ~{\rm GeV}$. (d) Same as (c) for the distribution in the transverse momentum of the di-lepton system. For all plots, acceptance cuts are applied (see fourth column in Tab.~\ref{tab:acceptance}).}}
\label{fig:numerics-SSM enhanced}
\end{center}
\end{figure}

We then analyse the SSM enhanced model. For this case, we assume an
integrated luminosity $L = 300$ fb$^{-1}$ corresponding to the value
expected at the LHC Run 3. Fig.~\ref{fig:numerics-SSM enhanced}a
displays in the main plot the invariant mass spectrum for the
SSM enhanced scenario with $M_{Z^\prime}=5$ TeV compared to the SM. We
plot both the LO and NNLL results. The QCD corrections increase
with the di-lepton invariant mass, reaching a 50$\%$ magnitude over
the LO result at the $Z'$ boson peak. The statistical
uncertainties are represented by the error bands on the NNLL results. The presence of a
wide $Z'$ boson in this case causes a ``shoulder'' in the
invariant mass spectrum beginning at around 3 TeV and continuing beyond
5 TeV.
%In particular, this shoulder lies outside the statistical uncertainty bands from before %4 TeV up through to $M_{Z^\prime}=5$ TeV and so may allow detection of this model below the %actual value of the $Z'$ boson mass. 
The potential excess of events as compared to the SM background is
clearly evidenced in the BSM/SM ratio presented in the sub-plot, where
the value of the ratio raises from 50 to 400 in the $4000~ {\rm GeV}\le
M_{ll}\le 5000~ {\rm GeV}$ mass window. However, in reality, for the
projected luminosity at the LHC Run 3 the total number of events that one
could count starting from a lower invariant mass threshold of 3 TeV is
fairly small. On top of that, the width is so large that no
discernible structure assignable to a $Z^\prime$ may exist in the
cross section mapped in the invariant mass of the di-lepton pair. Due
to the extreme lack of statistics and of any truly observable resonant
peaking structure, no strong claim could then be made on the existence
of new physics. We therefore consider other possible observables that
may have distinctive trends that are significantly
different from SM expectations, which may be exploited to establish a
signal in neutral DY final states. For a wide
$Z'$ boson, interference effects between the BSM signal and SM
background in such channels are significant. This implies two
key aspects of the associated phenomenology.
%several edits in text below
Firstly, unlike in the case of a narrow resonance where the BW peak
position, giving the $Z'$ boson mass, can readily be identified
in a model-independent way, the broad structure in the
$d\sigma/dM_{ll}$ distribution can no longer be optimally sought
by assuming a narrow resonance signal structure.  Secondly, other
spectra such as $dA_{\rm FB}/dM_{ll}$, $d\sigma/dp_T^{\rm min}$ or
$d\sigma/dq_T$ may show the aforementioned distinctive features away
from the $Z^\prime$ peak itself, including in the low invariant mass
tail where one would naively expect the SM to dominate.  While it
becomes impossible to design a model independent experimental search
giving the best possible sensitivity in all cases, it conversely
becomes possible to readily identify the underlying theoretical BSM
scenario in presence of a signal.

We first consider $A_{\rm FB}$, which has
the advantage that it is constructed from a ratio of cross sections
($d\sigma/dM_{ll}$) and hence benefits from the cancellation of both
experimental and theoretical systematic effects, as already intimated. Conversely, the
statistical error is much larger for $A_{\rm FB}$ than it is for
$d\sigma/dM_{ll}$.  Hence, the relative advantages of these
observables depends on the amount of integrated luminosity available at the
LHC. Fig.~\ref{fig:numerics-SSM enhanced}b shows the $A_{\rm{FB}}$
observable in the SSM enhanced scenario.  Once more the experimental statistical
error bands  dominate over the theoretical sources given by MC
errors and scale dependence. The latter are therefore not shown.
%We notice also that the NNLL result varies very little compared to the Born level
%prediction across the entire invariant mass range.
%, clearly demonstrating the robustness of the $A_{\rm{FB}}$ against the resummation of the
%transverse momentum logarithms. We therefore show only the NNLL $A_{\rm{FB}}$. 
Again both the stability of the $A_{\rm FB}$ observable with respect to higher orders and its complementarity to the invariant
mass spectrum are clear. For the forward-backward asymmetry the $Z'$ boson contribution, in fact,
deviates from SM expectations at a level greater than the expected experimental
statistical errors at invariant masses much lower than those where the ``shoulder'' starts
to emerge over the SM di-lepton mass spectrum. Here, considering the region
just above $M_{ll}\ge 1$ TeV where the $A_{\rm{FB}}$ starts showing the
effect of the presence of a $Z'$ boson, the number of events is
much larger than in the region in the mass spectrum where departures from the
SM expectation are observable, having of order one thousand events. Moreover, the $A_{\rm{FB}}$
shows a very well defined structure that significantly deviates from the
SM yield. In the case of such a wide $Z'$ boson, the measurement
of $A_{\rm FB}$ could be decisive for a discovery.

To support any preliminary evidence, one can consider other
observables to tackle the search. One could consider the distribution
in the minimum transverse momentum of a single lepton, $p_T^{\rm min}$, as shown in Fig.~\ref{fig:numerics-SSM enhanced}c. Here
we observe the relic of the Jacobian peak, flattened by the fact that
the $Z'$ boson is quite wide in this model. This peak is however
more pronounced than the deviation in the falling cross section of the
di-lepton mass spectrum, thus potentially helping to estimate the mass
$M_{Z^\prime}$.  A third variable of interest is the differential
cross section in the transverse momentum of the di-lepton pair,
$q_T$. We see that in a search window around the $Z'$ boson
mass, $4000 ~{\rm GeV}\le M_{ll}\le 6000 ~{\rm GeV}$ shown in Fig.~\ref{fig:numerics-SSM enhanced}d, the $q_T$
distribution is enhanced by the presence of the hypothetical
$Z'$ boson by a factor of one hundred or more compared to the SM
background. This effect is concentrated in the low $q_T$ range, as
expected (see also Fig.~\ref{qTspectrum_Z'vsSM_SSMwide}), and is statistically significant. The
measurement of $q_T$ could therefore support the observation of an
excess of events (a few) in the di-lepton spectrum and of a deviation
(sizeable) in the shape of the $A_{\rm{FB}}$, strengthening the experimental evidence  
in the event of the presence of new physics.

\begin{figure}[t]
\begin{center}
\includegraphics[width=0.5\textwidth]{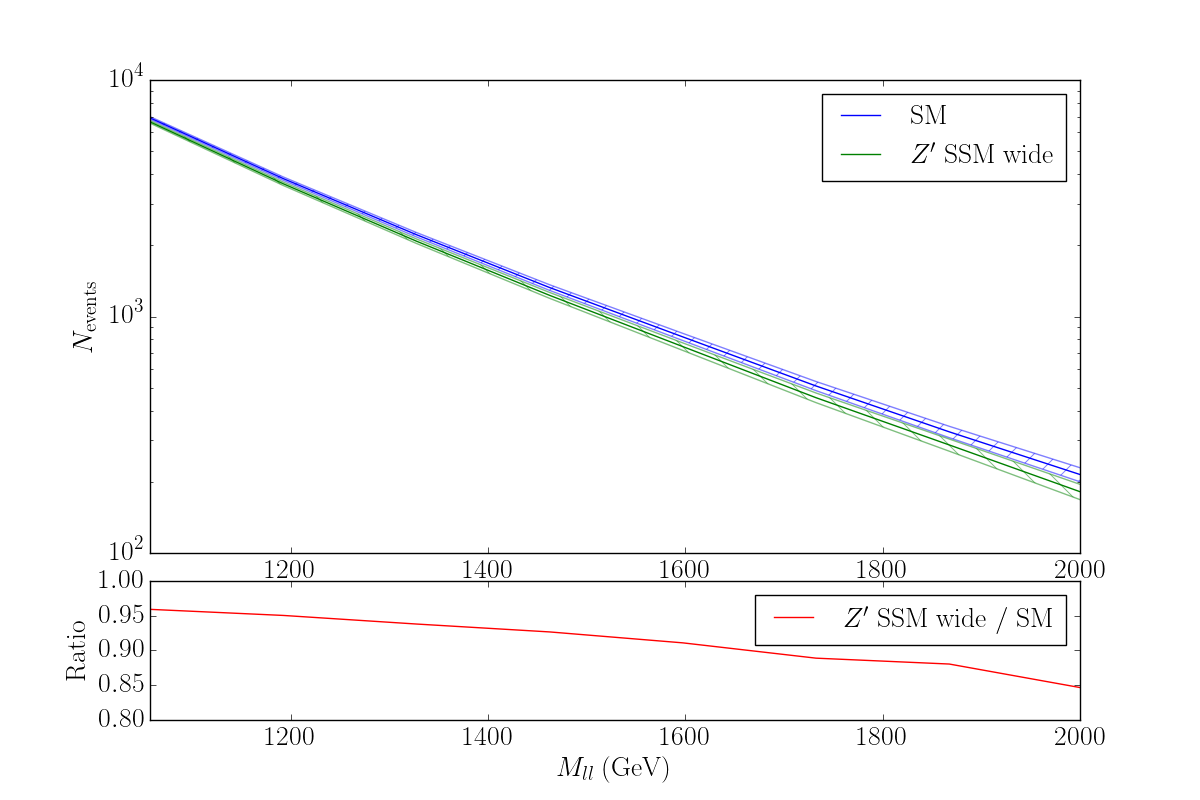}\includegraphics[width=0.5\textwidth]{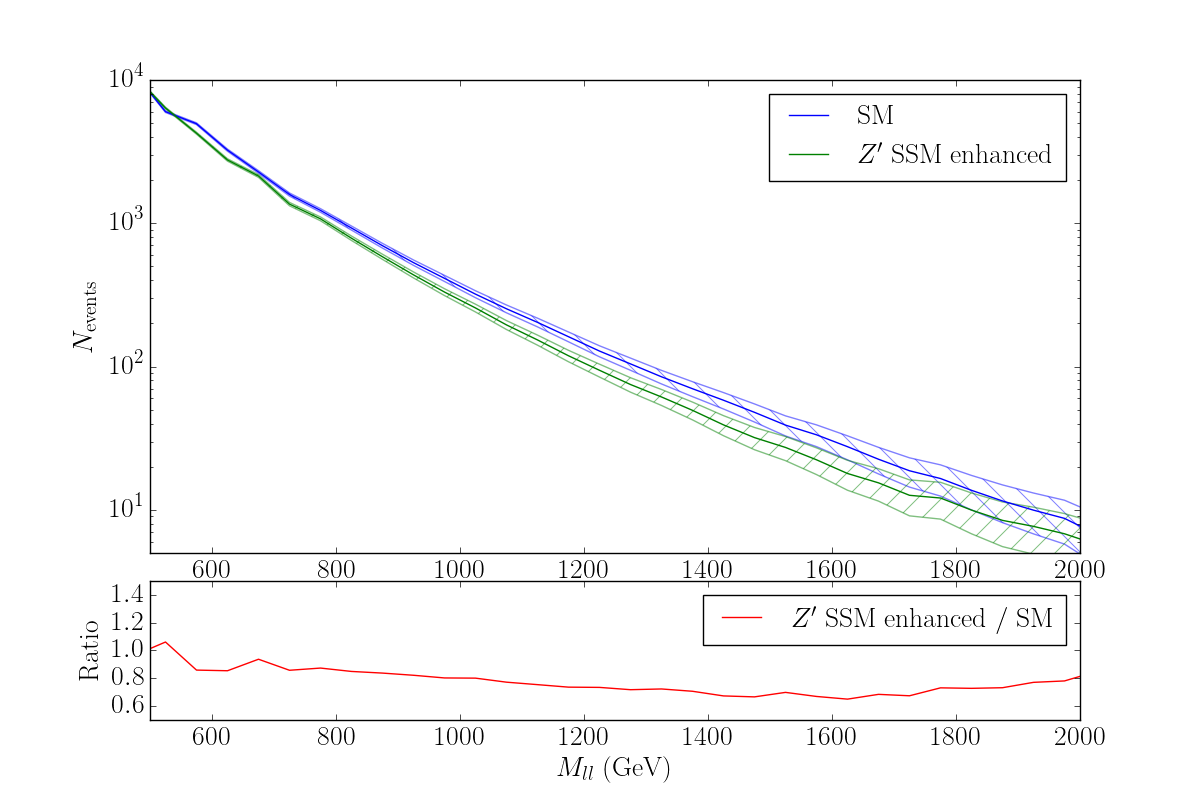}
\caption{\small{Zoom-in of the di-lepton spectrum (calculated at NNLL) in the low invariant mass range end for (a) SSM wide scenario with luminosity of $L=3000$fb$^{-1}$ and (b) SSM enhanced scenario with a luminosity of $L=300$ fb$^{-1}$, both compared to the SM. The statistical error bands are shown. The ratios of BSM/SM are given in the sub-plots.}} 
\label{Nevents_NNLL_ZpvsSM_SSMenhanced_zoomeddepletion}
\end{center}
\end{figure}

In addition to the excess of events predicted in both scenarios,
either around the $Z^\prime$ resonance peak for the SSM wide case or
spread over the shoulder for the SSM enhanced model, we also see a
depletion of events at lower invariant masses due to the interference
of the $Z'$ boson contribution with the SM $\gamma,Z$ one. This
is shown in Figs.~\ref{fig:numerics-SSM wide}a and
\ref{fig:numerics-SSM enhanced}a, however it is unclear due to the larger ranges and so we show zoomed-in versions focusing on the lower invariant mass portions in Fig.~\ref{Nevents_NNLL_ZpvsSM_SSMenhanced_zoomeddepletion}. In both scenarios the depletion of events in the $Z'$ case relative to the SM is statistically significant over at least part of this range. In the SSM enhanced case of Fig.~\ref{Nevents_NNLL_ZpvsSM_SSMenhanced_zoomeddepletion}b this effect is greater and extends over a wider invariant mass range and so is more promising. In this case a
depletion of about 20\% of the events for the SSM enhanced scenario considered could in principle be observed
in the $1000 ~{\rm GeV}\le M_{ll}\le 1500 ~{\rm GeV}$ mass window. In
conjunction with the shape difference seen at relatively low invariant
masses in the $A_{\rm{FB}}$, these effects at considerably lower invariant
masses than the resonance peak may offer a further means of probing
 $Z^\prime$ physics.

\section{Conclusions}

The  $A_{\rm FB}$ observable in $Z^\prime$ physics, other than being a time-honoured diagnostic probe, has recently been established to also be a discovery tool at the LHC whenever the
new neutral massive gauge boson is wide, i.e., it displays a large  ratio between its width  $\Gamma_{Z'}$ and  mass $M_{Z'}$. The advantages in this respect are twofold. Firstly, $A_{\rm FB}$ may reveal experimentally non-trivial structures in the invariant mass even when the cross section loses altogether its striking BW appearance. Secondly, it is much more stable than the latter in relation to systematic uncertainties from both the experimental and theoretical side, owing to $A_{\rm FB}$ being a ratio of cross sections. The first feature has been proven to be true quantitatively for a variety of $Z'$ models whereas the second one has recently been demonstrated for the case of the PDF error. Herein, we have complemented this last result by proving the stability of the $A_{\rm FB}$ also against higher-order effects entering the hard scattering, in the form of the leading resummed QCD perturbative corrections. Indeed, since it is most often the case that $A_{\rm FB}$ ought to be combined with cross section ($\sigma$) measurements in order to both achieve $Z'$ discovery and successfully diagnose its properties, we have extended the treatment of such higher-order QCD  effects also to the case of differential $\sigma$ observables. Crucially, this ought to be done not only for di-lepton quantities, but also for  single lepton ones, in order to ensure that the acceptance region of the detector is not returning different efficiencies in the presence of a wide $Z'$ (or these can be corrected for), with respect to the SM case. 
Here, we have studied all this using two benchmark scenarios, so-called `SSM wide' and `SSM enhanced', wherein the $Z'$ is always sufficiently wide
that sensitivity to it, above and beyond the SM yield,  may first emerge in the low-mass tail of the di-lepton mass distribution, rather than the peak region, no matter its shape (being indeed more BW-like in the former than in the latter case). As the  
potential observation of such 
phenomenological effects  is more  dependent on  experimental statistical  than   systematic errors, we have constructed these two scenarios in such a way that one (SSM enhanced) may be accessible at Run 3 luminosities while the other (SSM wide) may be so only with HL-LHC data samples. Finally, we have verified that the residual theoretical systematic error associated with the NNLL accuracy of our results is always much smaller in comparison to the experimental ones. 
% just like MC numerical errors are throughout.    
%
In short, at the LHC, wide $Z'$ scenarios may, on the one hand, no longer be elusive, thanks to a combination of $\sigma$ and $A_{\rm FB}$ measurements, and, on the other hand, be separable from one another, thanks to the stability of the distributions enabling such a separation against the dominant higher-order QCD effects.

\section*{Acknowledgements}
\noindent
This work is supported  by the Science and Technology Facilities Council, grant number  ST/L000296/1.
EA and SM acknowledge partial financial support through the NExT Institute. TC and CV thank the Science and Technology Facilities Council (STFC) for support via ST/P000274/1 and ST/R504671/1, respectively.
JF work has been supported by the BMBF under contract 05H15PMCCA and the DFG through the Research Training Network 2149 ``Strong and weak interactions - from hadrons to dark matter".  FH thanks DESY, Hamburg for hospitality and support while part of this work was being done.

%%%%%%%%%%%%%%%%%%%%%%%%%%%%%%%%%%%%%%%%%%%%%%%%%%%%%%%%%%%%%%%%%%%%%%%%%%%%%%%%

%%%%%%%%%%%%%%%%%%%%%

\end{document}